\documentclass[pre,aps,showpacs,byrevtex,amsmath,amssymb,preprint]{revtex4}

\usepackage{times}
\usepackage{epsf}
\usepackage{graphicx}

\usepackage{dcolumn}
\usepackage{bm}

\begin{document}
\draft

\title{Magnetophoresis of ferrofluid in microchannel system and its nonlinear effect}

\author{Y. C. Jian$^{1,2}$, L. F. Zhang$^1$ and
J. P. Huang$^1$\footnote{Electronic address: jphuang@fudan.edu.cn}}
\address{$^1$Surface Physics Laboratory (National Key Laboratory)
and Department of Physics, Fudan University, Shanghai 200433\\
$^2$Department of Physics, National Tsing-Hua University,
Hsinchu300}

\begin{abstract}
We have studied the magnetophoretic particle separation and its
nonlinear behavior of ferrofluids in microchannel which is proposed
by Furlani. The magnetic gradient force is caused by an bias field
and the polarized magnets and is found to be spatially uniform in
the channel section which can be used for particle selecting or
separation. We have derived the equations of nonlinear magnetization
of magnetic particles which cause the harmonics of magnetophoresis.
The Langevin model and generalized Clausius-Mossotti equation used
show how the normal and longitude anomalous anisotropic effect the
permeability of ferrofluids, thus the magnetic force. Our analysis
demonstrates the viability of using the microchannel system for
various bioapplications and other characterization of fluid
transporting and the time-varying magnetic field can be potentially
used for an integrated magnetometer and influences the the viscosity
and effective permeability in ferrofluids.
\end{abstract}


 \maketitle

\section{Introduction}

Nowadays magnetophoretic microsystems have been paid great attention
in bio-technology for the integration of
"micro-total-analysis"($\mu$TAS)~\cite{verpoorte} because of its
high degree of detection and selectivity. The biomaterials
possessing low magnetic susceptibility can cause substantial
contrast between the labeled and unlabled materials~\cite{furlani}.
Because of its polarization difference, the the particles exhibit
rich fluid-dynamic behaviors such as magnetophoresis as well as
various magnetic responses. Magnetic cell separation can be applied
using magnetic beads which coated with specific cell(core-shell
microsphere), or the native magnetic
susceptibility~\cite{bizdoaca,austin}. In some special cases such as
blood cells, the red and white blood cells can be conducted using
magnetophoretic separation based on their native magnetic
properties: diamagnetic or paramagnetic~\cite{han}. Magnetic changes
in red blood cells can also be used for separation of diseased
cells~\cite{paul}. In such continuous microseparator, the
ferromanetic wire(circular or square) put in close proximity under
an external bias field cause strong magnetic field with high
magnetic gradience. Miniaturized Cell separator can be integrated
for various types of cell counting and collecting. The
magnetophorsis with integrated soft magnetic elements have some
advantages over electrophoresis with electromagnets~\cite{choi} for
they consume no heat and cause no damage and other negative effect
on the bio-cells. Furlani have recently demonstrated transport and
capture behavior of magnetic particles such as Fe$_3$O$_4$ in the
microsystem which consists of an array of integrated soft magnetic
elements embedded under the microfluidic channel with also an
external bias field. The elements can be polarized by the bias
field, thus producing nonuniform field distribution which causes
magnetophoretic force on magnetic particles within microchannel. The
cubic soft magnetic elements replace the wire, producing different
separation and trap in geometry.

In the present paper we emphasize on the characteristics of the
magnetic composites(such as large magneticc constants and permanent
moments) on the behavior of magnetophoresis in ferrofluids and donot
consider the equations for particle motion. In experiment the
slow-moving transport is influenced by the viscous drag and thermal
kinetics, thus the magnetic force can be measured in the
quasi-equilibrium movement.

Ferrofluids(or magnetic fluids) are colloidal suspensions containing
single domain nanosized ferromagnetic particles dispersed in a
carrier fluid~\cite{skjeltorp}. Since these particles can interact
easily and form crystal-like structure in the presence of applied
bias magnetic fields, which in turn can affect the viscosity and
structural properties tremendously~\cite{huang}, particles in
ferrofluids have a wide variety of potential biomedical applications
such as label and manipulate biomaterials. The dynamic(ac) magnetic
properties and magnetization-induced second-order harmonic
generation are taken into consideration in the system. In
experiment, the measurement of an ac complex magnetic susceptibility
of magnetic fluids is a suitable method to study the relaxation
process of the magnetic dipoles of colloidal particles in magnetic
fluids~\cite{zhang, valenzuela}. The second-order harmonic
generation is the phenomenon that the magnetization along the
specific direction activates the originally silent tensor components
for the second-order nonlinear optical susceptibility and is
observed for surfaces or interfaces of ferromagnetic materials(thin
films)~\cite{bennemann}, or polar antiferromagnets~\cite{frohlich}
and polar ferromagnets~\cite{ogawa}.

The saturation of magnetic particles considered here will be
different from Furlani's theory and Han $et$ $al$'s experiment,
which cause the nonlinearity to appear by two effects: normal
saturation and anomalous saturation. In detail, the normal
saturation arises from the higher orders of Langevin function at
large filed region, and the anomalous saturation results from the
particle chains with higher and lower dipole moments~\cite{bottcher}
which is shifted under the influence of the field. The magnetic
field inside the ferrofluid plays important role in the coupling
between the two effects, which is similar in electrorheological and
magnetorheological fluids. When suspension having nonlinear
characteristic is subjected to ac magnetic field, the harmonics of
magnetic susceptibility can be induced to appear~\cite{levy}. Our
analysis demonstrates that the magnetic force will cause suitable
separation across microchannel and will be affected by the
anisotropic changes of magnetic dipole arrays.

\newpage
\section{Theory and Formalism}\label{}

\subsection{Nonlinear magnetic moments in ferrofluids}\label{}

In the standard case the magnetic induction $\textbf{B}$ is
proportional with the field strength $\textbf{H}_0$, and have the
relation $\textbf{B}=\mu\textbf{H}_0$, where $\mu$ is the linear
permeability. At strong field intensities, nonlinearities are
introduced as ${\bf B}=\mu_e {\bf H}_0+\chi H_0^2{\bf H}_0$ where
third-order and higher-order nonlinear coefficients are dropped and
$\chi$ and $\mu_e$ are the nonlinear magnetic susceptibility and
effective permeability for the longitudinal field case. Here we
assume the nonlinearity is not strong and consider only the
lowest-order nonlinearity for simplicity.

In the ferrofluids, the average component in the direction of the
field of the magnetic dipole moments can be expressed as $\langle
{\bf M}_d\cdot {\bf e}\rangle$, where $\langle\cdots \rangle$
denotes $\langle\cdots \rangle = \int \frac{\int {\rm d}X^{n_1-i}
\cdots e^{-u/k_BT}}{\int {\rm d}X^{n_1} e^{-u/k_BT}} {\rm d}X^i$,
${\bf e}$ is the unit vector in the direction of the external field,
and $X$ stands for the set of position and orientation variables of
all particles. Here $u$ is the energy related to the dipoles in the
sphere, and it consists of three parts: the energy of the dipoles in
the external field $u_{\rm de}$, the magnetostatic interaction
energy of the dipoles $u_{\rm mi}$, the non-magnetostatic
interaction energy between the dipoles $u_{\rm nmi}$ which is
responsible for the short-range correlation between orientations and
positions of the dipoles such as London-Van der Waals interaction
energy.

In this case, the effective permeability of ferrofluid $\mu_e$ is
determined by the generalized Clausius-Mossotti equation taking into
consideration of dipolar interactions~\cite{lo,bottcher}:

\begin{equation}
\frac{g_L(\mu_e-\mu_2)}{\mu_2+g_L(\mu_e-\mu_2)} = \frac{N}{3}\left
(\alpha+ \frac{\langle M_d^2\rangle}{3k_BT}\frac{1}{1+i 2\pi
f\tau}\right ),
\end{equation}

where $\mu_2$ represents the permeability of the host fluid, $N$ the
number density of the particles, $k_B$ the Boltzmann constant, $T$
the absolute temperature, $f$ the frequency of the applied magnetic
field, $\tau$ the relaxation time of the particles, and $\alpha$ the
magnetizability of the particles. Here $\tau$ can be expressed as
$\tau_b\tau_n/(\tau_b+\tau_n)$, where $\tau_b$ and $\tau_n$ are the
brownian relaxation time and the n$\rm\acute{e}el$ relaxation time
respectively~\cite{shliomis}. Our model describes the aggregation
behavior in an external field by introducing the longitudinal
demagnetizing factor $g_L$ in Clausius-Mossotti equation, which
deserves a thorough consideration: Eq. (2) should be expected to
contain both $u_{\rm mi}$ and $u_{\rm nmi}$ as $g_L$ is not equal to
$1/3$. For an isotropic array of magnetic dipoles, the demagnetizing
factor will be diagonal with the diagonal element $g_L=1/3.$
However, in an anisotropic array like ferrofluid, the demagnetizing
factor can still be diagonal, but it deviates from $1/3$. In fact,
the degree of anisotropy of the system is just measured by how $g_L$
is deviated from $1/3.$ It is worth noting that $g_L\le 1/3$ in the
present longitudinal field case. Furthermore, there is a sum rule
for the factors, $g_L+2g_T =1$~\cite{landau}, where $g_T$ denotes
the transverse demagnetizing factor. Such factors were measured by
means of computer simulations~\cite{martin}. Thus, to investigate
the anisotropic structural information of the array,  we have to
modify the Clausius-Mossotti equation accordingly by including the
demagnetizing factor. When we studied the field-induced structure
transformation in ferrofluids, we can use the generalized
Clausius-Mossotti equation~\cite{lo} by introducing a local-field
factor $\beta'$ which reflects the particle-particle interaction
between the particles in a lattice.

The magnetic dipole moment ${\bf m}_i$ satisfies Langevin function
${\rm m}_i={\rm m}_s({\rm coth}\gamma-\frac{1}{\gamma})$ where
$\gamma = {\rm m}_0 H_0/(k_BT)$ and ${\bf m}_s$ denotes the
saturation magnetization of particles. For the whole array of
magnetic moments, $ {\bf M}_d = \sum_{i=1}^{n_1}({\bf
m}_i(\mu_{\infty}+2\mu_2)/3\mu_2)_i$ where $\mu_{\infty}$ represents
the permeability at frequencies at which the permanent dipoles
cannot follow the changes of the field but where the atomic and the
electronic magnetization are still the same as in the static
field~\cite{bottcher}. Therefore, $\mu_{\infty}$ is the permeability
characteristic for the induced magnetization. In practice,
$\mu_{\infty}$ can be expressed in the expression containing an
intrinsic dispersion,

\begin{equation}
\mu_{\infty} = \mu_{\infty}(0)+\frac{\Delta\mu}{1+i f/f_c}
\end{equation}

where $\mu_{\infty}(0)$ is the high-frequency limit permeability,
and $\Delta\mu$ stands for the magnetic dispersion strength with a
characteristic frequency $f_c$. Harmonics of magnetic moments can be
obtained through Fr$\rm\ddot{o}$hlich model~\cite{frohlich2} by

\begin{equation}
{\bf M}_d\cdot {\bf e}=-\frac{\partial u}{\partial H_F}.
\end{equation}

Here $H_F$ gives the magnetic field inside the spherical situated in
medium with permeability $\mu_e$, and has the form
$\frac{3\mu_e}{2\mu_e+\mu_{\infty}}H_0+\frac{3\chi
\mu_{\infty}}{(2\mu_e+\mu_{\infty})^2}H_0^3$. Thus taking into
account the higher derivatives of the average moment we obtain

\begin{eqnarray}
\langle {\bf M}_d\cdot {\bf e}\rangle=\frac{\partial\langle {\bf
M}_d\cdot {\bf e}\rangle }{\partial
H_F}|_{H_F=0}H_F+\frac{1}{6}\frac{\partial^3\langle {\bf M}_d\cdot
{\bf e}\rangle }{\partial H_F^3}|_{H_F=0}H_F^3.
\end{eqnarray}

Noticing the expression for $\langle {\bf M}_d\cdot {\bf e}\rangle$
and Eq. (3), it is easily derived

\begin{eqnarray}
\frac{\partial}{\partial H_F}\langle{\bf M}_d\cdot{\bf e}\rangle
|_{H_F=0}=\frac{1}{k_BT}[\langle({\bf M}_d\cdot{\bf e})^2\rangle-
\langle{\bf M}_d\cdot{\bf
e}\rangle^2]|_{H_F=0}&=&\frac{1}{k_BT}\langle M_d^2\rangle _0,
\end{eqnarray}
\begin{eqnarray}
\nonumber\\
& &\frac{\partial^3}{\partial H_F^3}\langle {\bf M}_d\cdot {\bf
e}\rangle |_{H_F=0}=\frac{1}{(k_BT)^3}[\langle ({\bf M}_d\cdot {\bf
e})^4\rangle -3\langle ({\bf M}_d\cdot {\bf
e})^3\rangle\langle  {\bf M}_d\cdot {\bf e}\rangle \nonumber\\
     & &                                    +6\langle ({\bf
M}_d\cdot {\bf e})^2\rangle \langle {\bf M}_d\cdot {\bf e}\rangle
^2  -3\langle ({\bf M}_d\cdot{\bf e})^2\rangle^2\nonumber\\
     & &                                    -6\langle {\bf
M}_d\cdot {\bf e}\rangle ^4+6\langle {\bf M}_d\cdot {\bf e}\rangle^2
\langle ({\bf M}_d\cdot {\bf
e})^2\rangle \nonumber\\
     & &                                    -\langle {\bf M}_d\cdot {\bf e}\rangle \langle  ({\bf
M}_d\cdot {\bf e})^3\rangle]|_{H_F=0}\nonumber\\
     & &                                    =\frac{1}{15(k_BT)^3}[3\langle M_d^4\rangle _0-5\langle
M_d^2\rangle ^2_0].
\end{eqnarray}

Using the same method, we obtain
\begin{eqnarray} \frac{\langle M_d^2\rangle
_0}{V}&=&\left(\frac{\mu_{\infty}+2\mu_2}{3\mu_2}\right)^2[\frac{n_1}{V}
p_0^2\sum_{j=1}^{n_1}\langle \cos \theta_{ij}\rangle ], \\
\frac{\langle M_d^4\rangle
_0}{V}&=&\left(\frac{\mu_{\infty}+2\mu_2}{3\mu_2}\right)^4[\frac{n_1}{V}
p_0^4\sum_{j=1}^{n_1}\langle \cos
\theta_{ij}\sum_{r=1}^{n_1}\sum_{s=1}^{n_1}\cos \theta_{rs}\rangle
].
 \end{eqnarray}

For the numerical calculation, we have
\begin{eqnarray}
\sum_{j=1}^{n_2}\langle \cos \theta_{ij}\rangle &=&1 , \\
\sum_{j=1}^{n_2}\langle \cos
\theta_{ij}\sum_{r=1}^{n_2}\sum_{s=1}^{n_2}\cos \theta_{rs}\rangle
&=&\frac{1}{3}(5 n_1-2).
\end{eqnarray}

In the presence of external oscillating time-varying magnetic
field~cite\cite{asbury}, the magnetic particles will have nonlinear
characteristics. In the experiment, the second term or higher order
terms of magnetization or force can be obtained using
$\emph{mixed-frequency measurements}$~\cite{yang}. When we apply an
external field such as $H_0(t)=H_{{\rm dc}}+H_{{\rm ac}}(t)=H_{{\rm
dc}}+H_{{\rm ac}}\sin \omega t$, the orientational magnetization
$M_{{\rm o}}$ will contain harmonics as
\begin{equation}
 M_{{\rm o}}=M_{{\rm
o}}^{{\rm (dc)}}+M_{\omega}\sin\omega t+M_{2\omega}\cos 2\omega t
+M_{3\omega}\sin 3\omega t+\cdots.
\end{equation}

Here $\omega=2\pi f $ and  $H_{{\rm dc}}$ denotes the dc field which
induces the anisotropic structure in the ferrofluids, and $H_{{\rm
ac}}(t)$ stands for a sinusoidal ac field. Applying Eq. (11) into
Eq. (4), (5) and (6), after tedious calculation, the harmonics of
magnetic moments can be expressed as

\begin{eqnarray}
M_{{\rm o}}^{{\rm (dc)}} &=&  H_{{\rm dc}}J_1+\frac{3}{2}H_{{\rm ac}}^2H_{{\rm dc}}\chi+H_{{\rm dc}}^3\chi,\\
M_{\omega} &=& H_{{\rm ac}}J_1+\frac{3}{4}H_{{\rm ac}}^3\chi+3H_{{\rm ac}}H_{{\rm dc}}^2\chi,\\
M_{2\omega} &=& -\frac{3}{2}H_{{\rm ac}}^2H_{{\rm dc}}\chi,\\
M_{3\omega} &=& -\frac{1}{4}H_{{\rm ac}}^3\chi,
\end{eqnarray}
with $J_1=\mu_e-\mu_{\infty}.$ Below we set $H_{{\rm ac}}=H_{{\rm
dc}}=H$ for simplicity, thus all harmonics for $M_0$ can all be
expressed as $M_0=\chi^{'}H$.

\subsection{Magnetophoresis in microchannel}\label{}

Now we will investigate the magnetophoretic behavior in
microchannel. In a standard case, a magnetically polarizable object
will be trapped in a region of a focused magnetic field, provided
there is sufficient magnetic response to overcome thermal energy and
the magnetophoretic force~\cite{jones}. For a magnetically linear
particle under magnetophoresis, the effective magnetic dipole moment
vector induced inside takes a form very similar to that for the
effective dipole moment of dielectric paricle, $\vec{m}=4\pi R_{\rm
eff}^{3} \vec{M_0}\rm Re[\rm CMF]$, where $R_{\rm eff}$ is the
effective radius of spherical or spheroidal particle and CMF is the
Clausius-Mossotti factor along the direction of external field. It
is noted that particles are attracted to magnetic field intensity
maxima when $b>0$ as positive magnetophoresis and negative
magnetophoresis corresponds to $b<0$. For spheroidal particle with
permeability $\mu_2$, $\rm
CMF=\frac{1}{3}\frac{\mu_2-\mu_e}{\mu_e+g_L(\mu_2-2\mu_e)}$. The
magnetophoretic force exerted on the particle in a nonuniform
magnetic field $\vec{H}$ can be written as,

\begin{equation}
\vec{F_m}=4\pi\mu_{e}R_{\rm eff}^{3} {\rm Re[CMF]}
\vec{M}_0\cdot\nabla \vec{H}
\end{equation}

The microsystem consists of one integrated soft-magnetic elements
which is embedded in a nonmagnetic substrate beneath a microfluidic
channel as shown in Fig. (1), in which the magnet is $2w$ wide and
$2h$ high. The magnetic particles in ferrofluids which pass through
the channel can separated according to the different field gradient
distribution, and the nonmagnetic particles will be rinsed away. Y
axe corresponds to the longitude case mentioned above. Below we will
show that when the number of magnetic element increases, the
magnetophoresis will become more uniform in both parallel and
perpendicular direction to that of particle transport. The bias
field $\vec{H}_{\rm bias}$ along y axe and field from the magnet
$\vec{H}_{\rm mag}$ which is saturated(saturated magnetization
$M_{\rm es}$) when the bias field is added, both contribute to the
magnetic field exerted on the particle. Under high frequency of bias
field($>10^3$Hz), the magnetization of magnet can be viewed
constant. Thus the magnetophoretic force is decomposed into two
components:

\begin{equation}
F_{\rm mx}=4\pi\mu_{e}R_{\rm eff}^{3} {\rm Re[CMF]}\chi^{'}[H_{\rm
mag,x}\frac{\partial H_{\rm mag,x}}{\partial x}+(H_{\rm
mag,y}+H_{\rm bias,y})\frac{\partial H_{\rm mag,y}}{\partial y}],
\end{equation}

\begin{equation}
F_{\rm my}=4\pi\mu_{e}R_{\rm eff}^{3} {\rm Re[CMF]}\chi^{'}[H_{\rm
mag,x}\frac{\partial H_{\rm mag,y}}{\partial x}+(H_{\rm
mag,y}+H_{\rm bias,y})\frac{\partial H_{\rm mag,x}}{\partial y}].
\end{equation}

where $\vec{H}_{\rm mag}$ can be expressed as $H_{\rm
mag,x}=\frac{M_{\rm es}}{4\pi}\{{\rm
ln}[\frac{(x+w)^2+(y-h)^2}{(x+w)^2+(y+h)^2}]-{\rm
ln}[\frac{(x-w)^2+(y-h)^2}{(x-w)^2+(y+h)^2}]\}$ and $H_{\rm
mag,y}=\frac{M_{\rm es}}{2\pi}\{{\rm
tan}^{-1}[\frac{2h(x+w)}{(x+w)^2+y^2-h^2}]-{\rm tan}^{-1}[
\frac{2h(x-w)}{(x-w)^2+y^2-h^2}]\}$ from ~\cite{furlani}. The
magnetophoretic force across the channel can then be be accurately
calculated. It could be noted from the equations above that without
the bias field magnet-element cannot be magnetized and bias field
contributes to the magnetophoretic force, but the bias field itself
produce no field gradient in the microchannel. Because the
importance of the the contribution from the bias field which causes
harmonics of magnetization in the particle, below we will calculate
it alone.

\section{Results}\label{}

Now we are in the position to study the behavior of magnetophoresis
in ferrofluids. Specifically, the external field is set to be
$H_{{\rm dc}}=H_{{\rm ac}}=0.01$A/m, high-frequency limit
permeability $\mu_{\infty}(0)=8\pi\times10^{-7}$H/m, dispersion
strength $\Delta\mu=32\pi\times10^{-7}$H/m, medium permeability
$\mu_2 =4\pi\times10^{-7}$H/m, permanent magnetic moment for each
particle $m_0=10^{-13}$A/m, characteristic frequency $f_c =5\times
10^3$Hz, the relaxation time $\tau=5\times 10^{-7}$s, the number
density $N =10^{5}$m$^{-3},$ the length of magnet $w=h=50\mu{\rm
m}$, Boltzmann constant $k_B = 1.38\times 10^{-23}$J/K, the
effective radius $R_{\rm eff}=500$nm and the average magnetization
of single particle $\alpha=10^{-5}$m$^{-3}$.

Figure 2 displays the temperature and anisotropic effect on CMF in
magnetophoresis. In Fig. 2(a), it is shown that increasing the
temperature T causes slight increase in CMF(about 3$\%$), thus has
little effect in magnetophoresis. From Fig. 2(b), the magneophoretic
force exhibits strong sensitivity to the anisotropic factor $g_L$.
When CMF is negative(particles will be repelled from the maximum
magnetic gradient), the larger $g_L$ is, the smaller repulsion will
be. When CMF is positive, with increaseing $g_L$ such an attractive
force becomes stronger. Furthmore, we predict the
anisotropic-dependent crossover frequencies at which there is no net
force on the cell particle. The crossover frequency is monotonically
increasing function of $g_L$, dependent on whether the variation of
magnetophoretic force is negative or positive.

Figure 3 and 4 shows the oriental, fundamental, second- and third-
order harmonics of the magnetic force $F_x$ and $F_y$ induced by the
bias field mentioned above as a function of the field frequency for
various $g_L$ in the longitudinal field case. The particle is put in
the position x=50nm and y=300nm(see the coordinate in the inset of
Fig. 5). The harmonics of magetization of particles will change
accordingly as the system alters from isotropic case ($g_L=1/3$) to
anisotropic ($g_L\ne 1/3$) because of the appearance of the particle
chains. In detail, stronger anisotropy (namely, decreasing the
longitudinal demagnizing factor $g_L$) leads to larger magnetic
force in the low-frequency region. It can also be observed there is
two plateaus in the range 0.03-3MHz and $10^2$-$10^4$MHz where the
magnetic force have slight change and the second and third order
harmonics of magnetic force are negligible compared with lower order
ones. In fact for the transverse field case, it could be concluded
from the sum rule between $g_L$ and $g_T$ that $g_L+2g_T=1.$ Because
of the coupling between the applied dc and ac magnetic fields, the
even-order harmonics are also induced to appear besides the
odd-order harmonics for the longitudinal field case, even though
only the cubic nonlinearity is considered due to the virtue of
symmetry of the system. In addition, the harmonics shown in Figs. 3
and 4 are nonzero at $g_L=1/3$ is because of the presence of
external fields (i.e., $u_{\rm de}\ne 0$), even though there is no
particle interactions (i.e., $u_{\rm mi}=u_{\rm nmi}=0$) as
$g_L=1/3$, the nonlinear behavior due to the normal saturation could
still be induced to occur.

Fig. 5 and 6 displays the magnetic force which is spatially varying
across the channel section, when one and two magnets are embedded in
the substrate under frequency 10KHz of magnetic field. In Fig. 6 the
two magnets are embedded 250$\mu$m away and the $x-y$ coordinate is
placed in the midpoint of magnets. Comparing Fig. 5(A) with Fig.
6(A), we find that the force $F_x$ for the purpose of separating
magnetic particles which is attractive in some regions and repulsive
in others along the $x$ axe becomes more uniform, and the force
$F_y$ has similar nature. In detail the absolute value of magnetic
force $F_x$ is symmetric in the $x$ direction and become stronger on
the edge of the microchannel, while $F_y$ is always negative which
will repel the particles from the magnets. The uniform separation
technique will cause the magnetic particles with different size or
permeability become apart into layers in geometry which  can in fact
not only be used in ferrofluids. The calculation also shows that
when more magnets are uniformly embedded, the separation of
particles are more efficient. It should be pointed out that the
distance between the nearby particles is important to cause
spatially uniform magnetic force and it can be chosen for different
purpose of separation which may be determined by the capture
efficiency for a specific particle sorting or transporting. The
time-varying magnetic field can be potentially used for an
integrated magnetometer and influences the the viscosity and
effective permeability in ferrofluids.

\section{Conclusion}\label{}

We have studied the magnetophoretic particle separation of
ferrofluids in microchannel and its nonlinear behavior. The magnetic
gradient force is caused by an bias field and the polarized magnets
and is found to be spatially uniform in the channel section which
can be used for particle selecting or separation. We have derived
the equations of nonlinear magnetization of magnetic particles which
cause the harmonics of magnetophoresis. The Langevin model and
generalized Clausius-Mossotti equation used show how the normal and
longitude anomalous anisotropic effect the permeability of
ferrofluids, thus the magnetic force. Our analysis demonstrates the
viability of using the microchannel system for various
bioapplications and other characterization of fluid transporting.

\section*{Acknowledgements}
Y.C.J is grateful to Prof. Chia-Fu Chou for the generous help and
hospitality at Sinica and Wu Ta-you Camp in Taiwan in the academic
year 2006 supported by ChunTsung(T. D. Lee) Foundation. The authors
thank Prof. T. Nakayama from Hokkaido University in Japan for great
support and acknowledge the financial support by the Shanghai
Education Committee and the Shanghai Education Development
Foundation (¡±Shu Guang¡± project) under Grant No. KBH1512203, by
the Scientific Research Foundation for the Returned Overseas Chinese
Scholars, State Education Ministry, China, by the National Natural
Science Foundation of China under Grant No. 10321003.

\newpage
\section*{Figure Captions}
Fig. 1. (Color online) Schematic graph showing one integrated
soft-magnetic elements embeds in a nonmagnetic substrate beneath a
microfluidic channel through which ferrofluids flow.

Fig. 2. (a) CMF for different temperatures vs the frequency f of
magnetic fields. (b) CMF vs the frequency fof magnetic fields for
different anisotropic factors $g_L$.

Fig. 3. Oriental, Fundamental, second and third order harmonics of
the magnetic force $F_x$ vs the field frequency for various $g_L$ in
the longitudinal field case.

Fig. 4. Same as Fig. 3, but for another force component $F_y$.

Fig. 5. (Color online) (A) The spatial magnetic force $F_x$ across
the channel section, when one magnet embeds in the substrate under
frequency 10KHz of magnetic field. (B) The spatial magnetic force
$F_y$ across the channel section.

Fig. 6. (Color online) Same as Fig. 5, but for two magnets case.

\newpage
\begin{figure}[h]
\includegraphics[width=200pt]{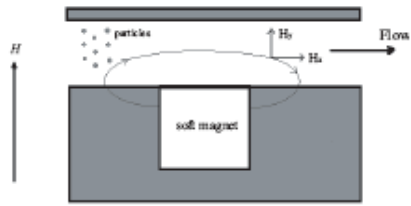}
\caption{Jian, Zhang and Huang}.\label{}
\end{figure}
\newpage
\begin{figure}[h]
\includegraphics[width=200pt]{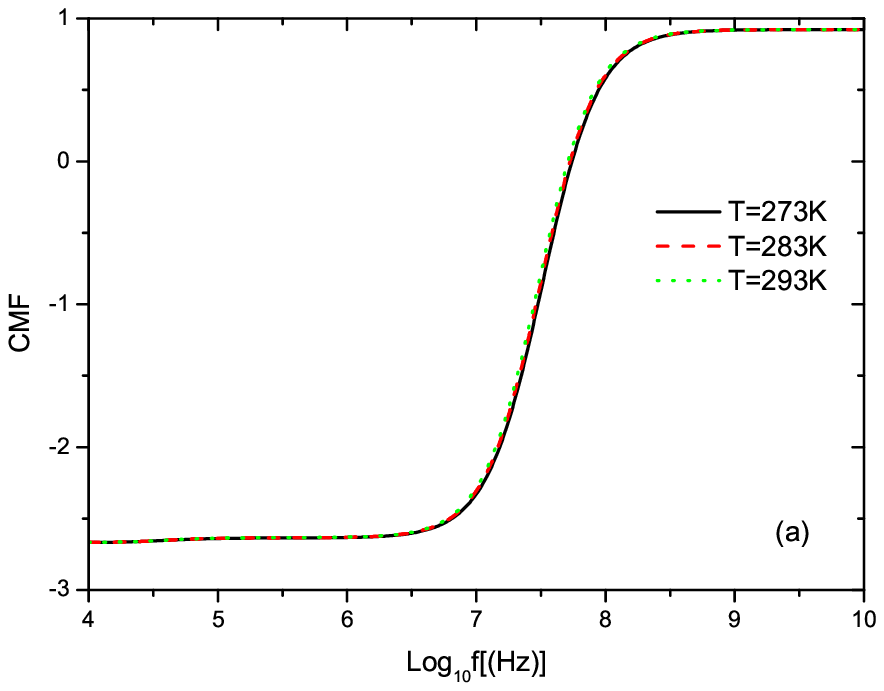}
\includegraphics[width=200pt]{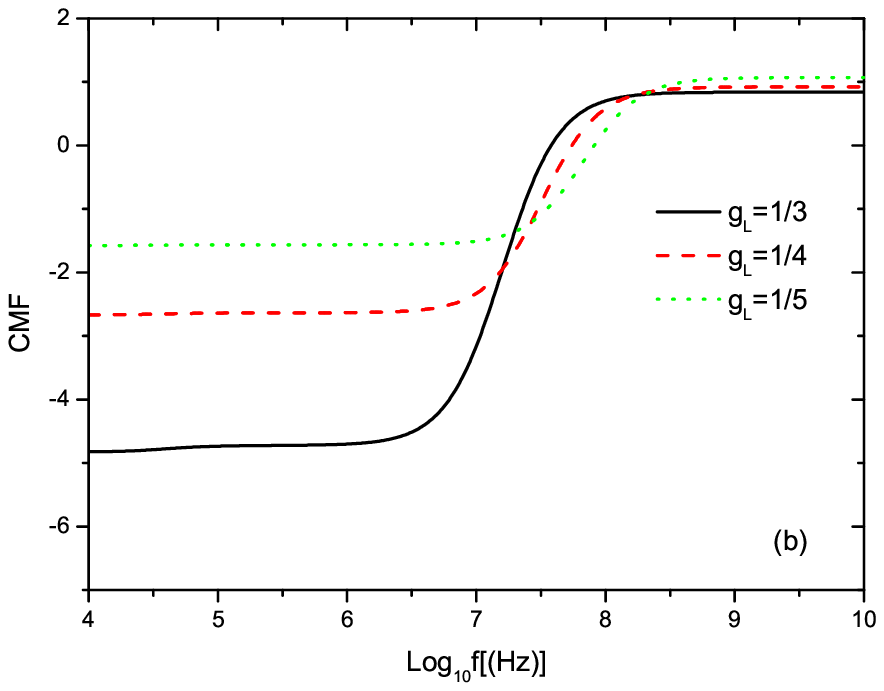}
\caption{Jian, Zhang and Huang}.\label{}
\end{figure}
\newpage
\begin{figure}[h]
\includegraphics[width=200pt]{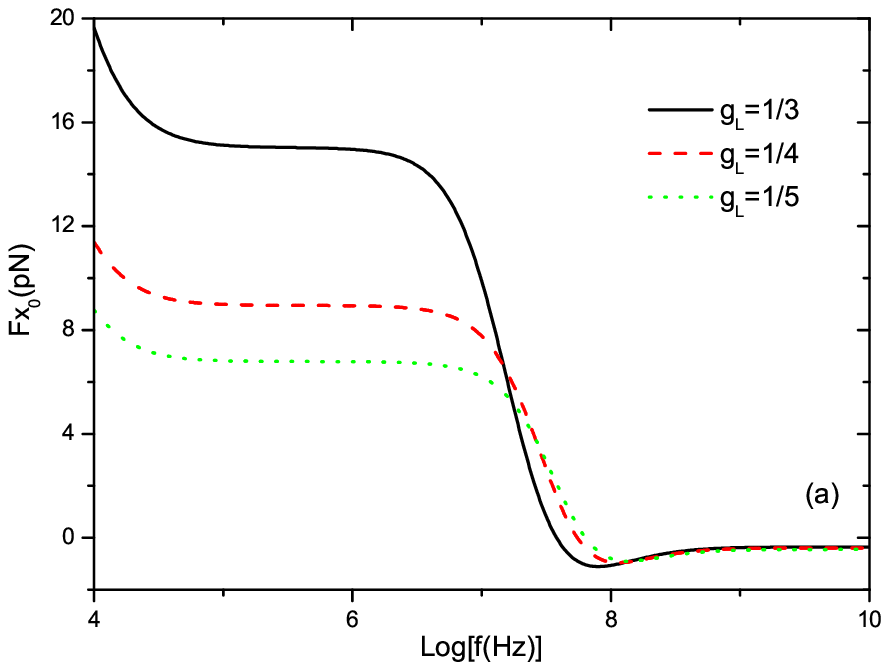}
\includegraphics[width=200pt]{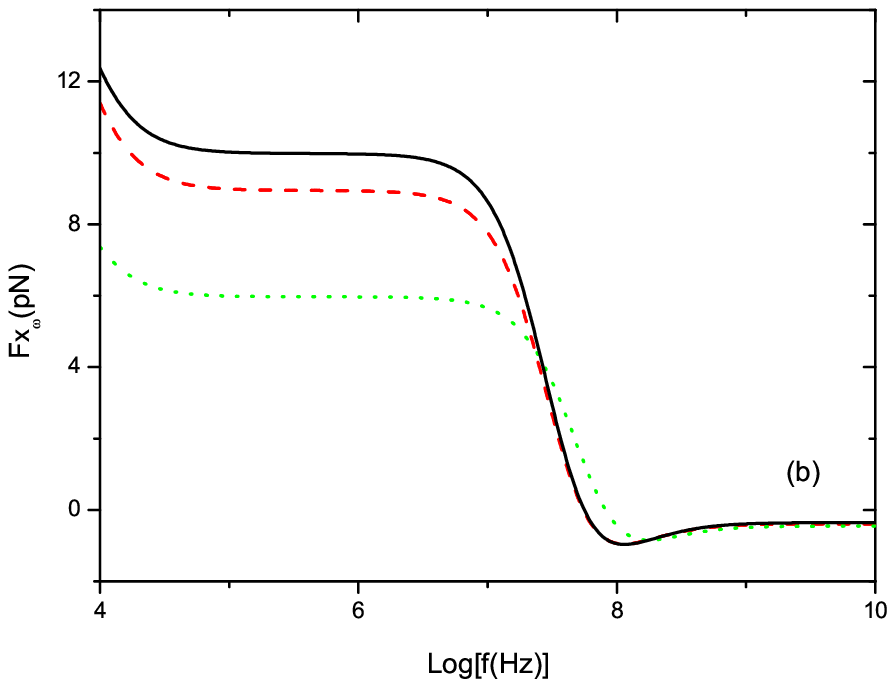}
\includegraphics[width=200pt]{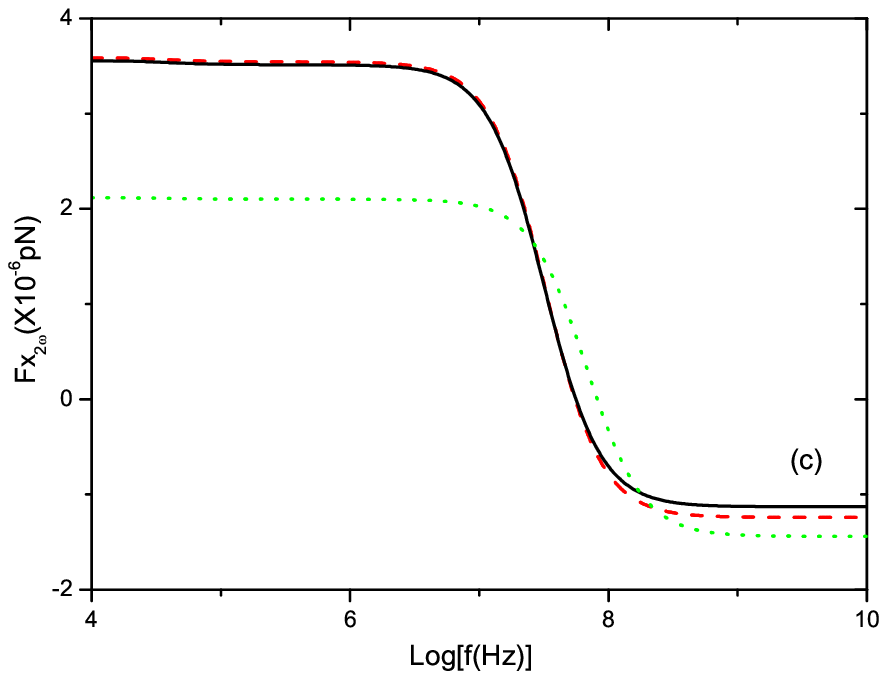}
\includegraphics[width=200pt]{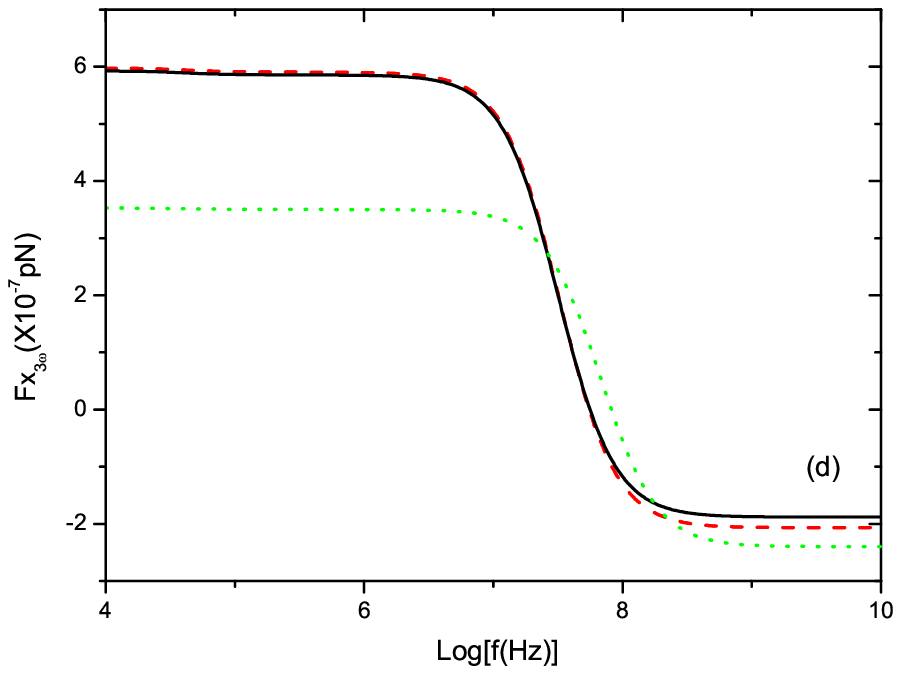}
\caption{Jian, Zhang and Huang}.\label{}
\end{figure}
\newpage
\begin{figure}[h]
\includegraphics[width=200pt]{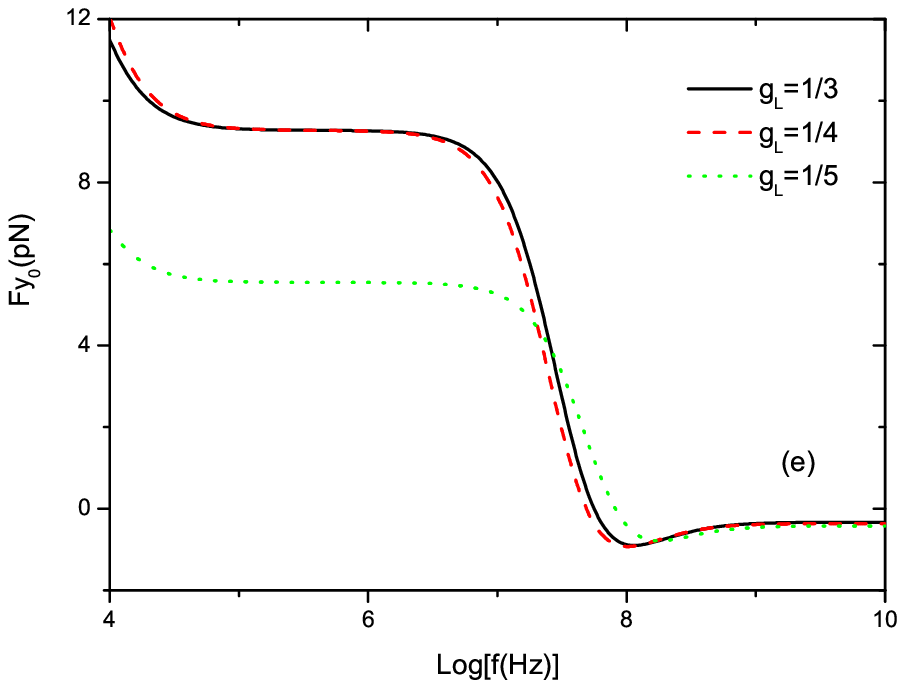}
\includegraphics[width=200pt]{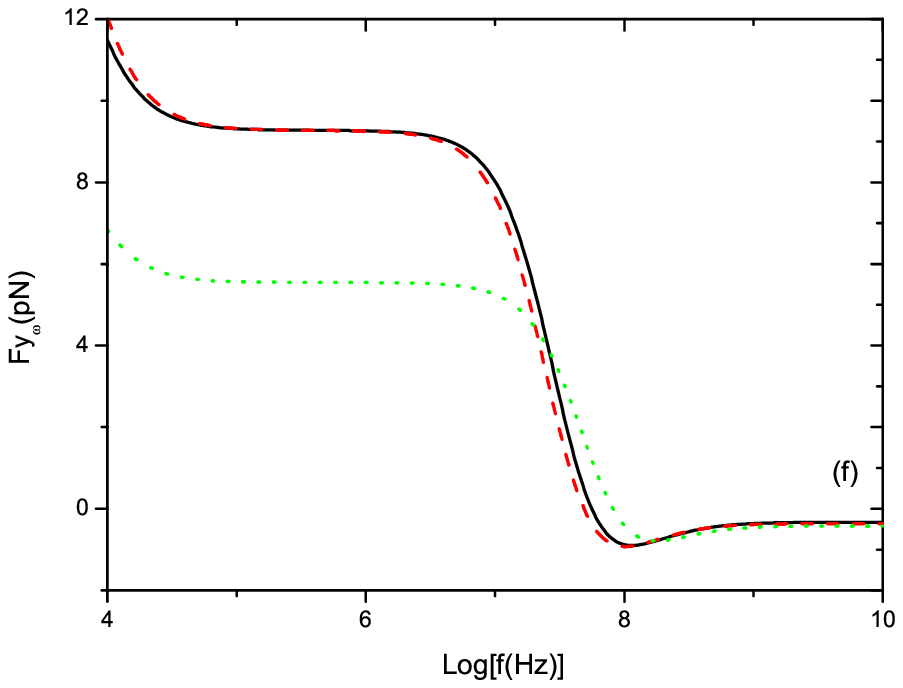}
\includegraphics[width=200pt]{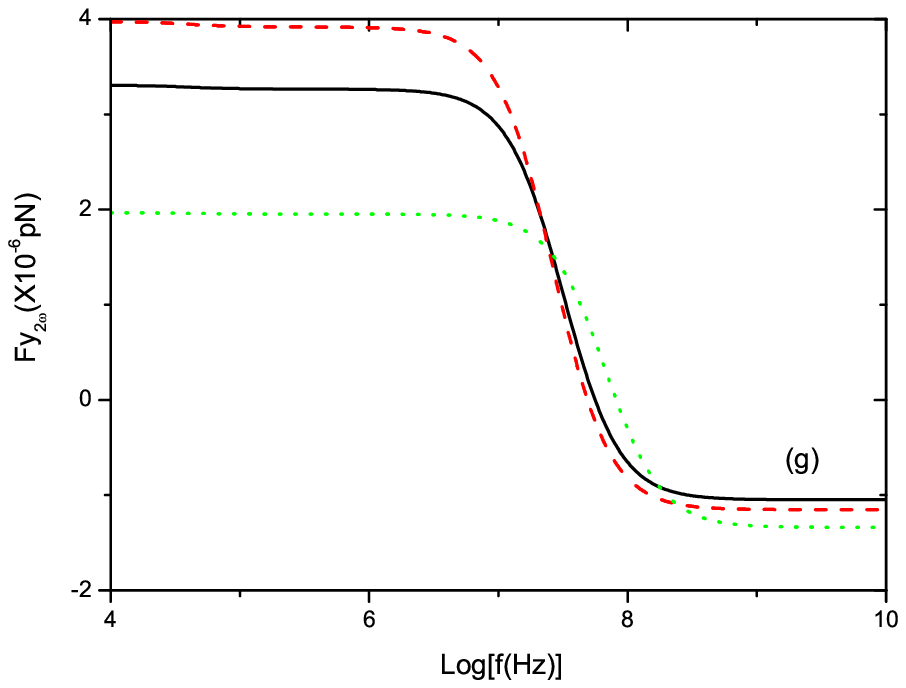}
\includegraphics[width=200pt]{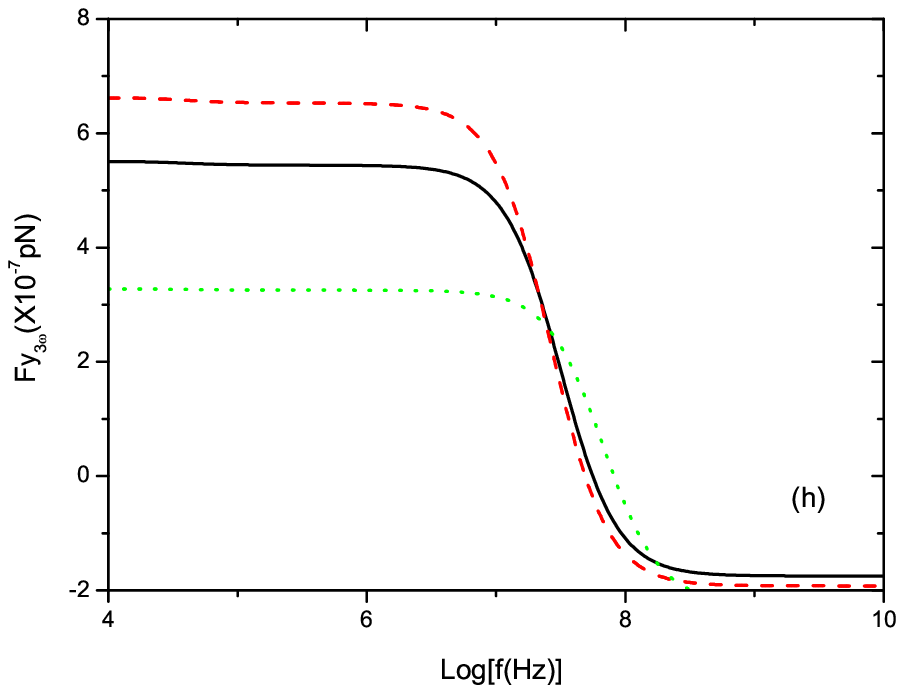}
\caption{Jian, Zhang and Huang}.\label{}
\end{figure}
\newpage
\begin{figure}[h]
\includegraphics[width=300pt]{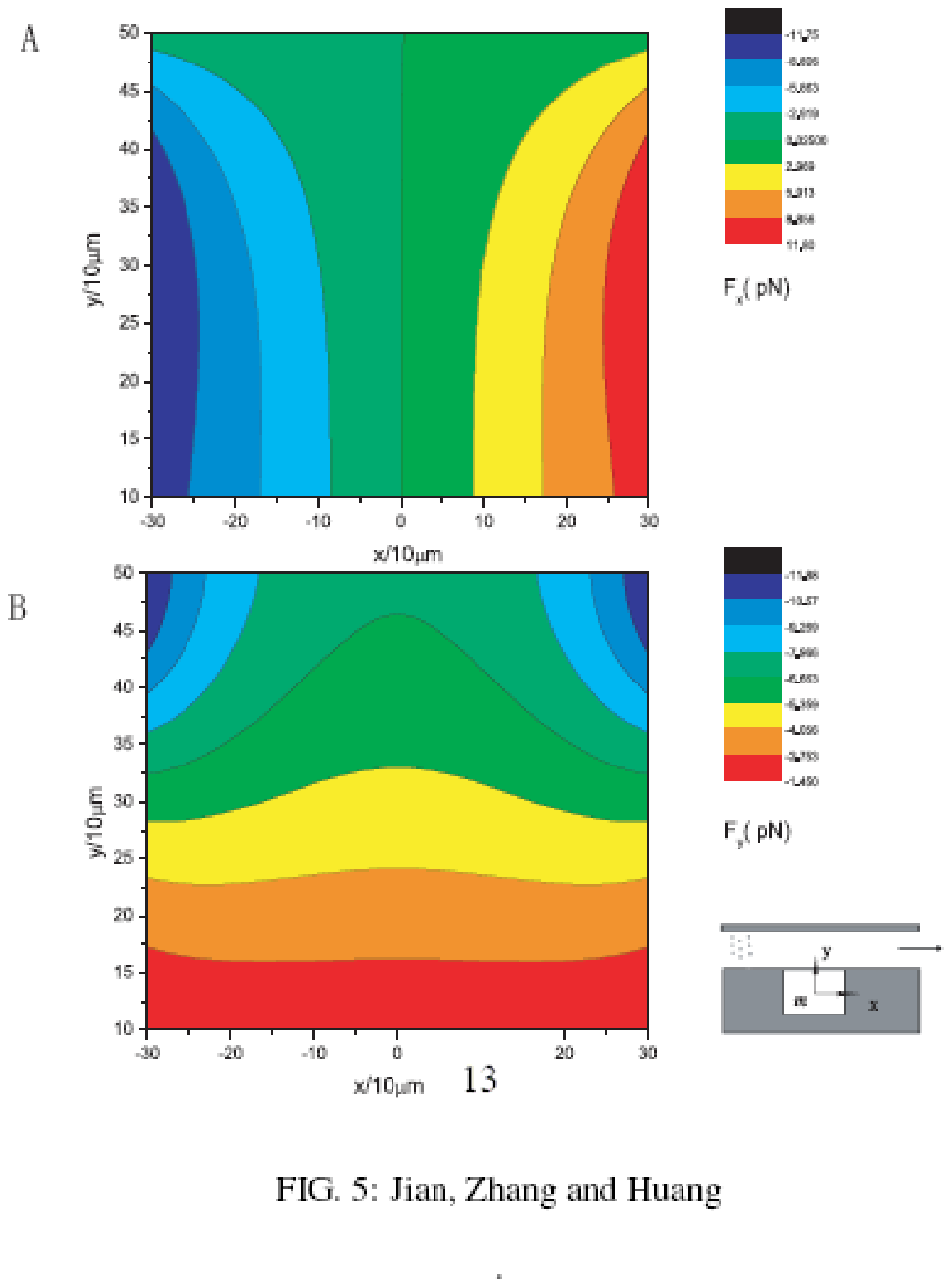}
\caption{Jian, Zhang and Huang}.\label{}
\end{figure}
\newpage
\begin{figure}[h]
\includegraphics[width=300pt]{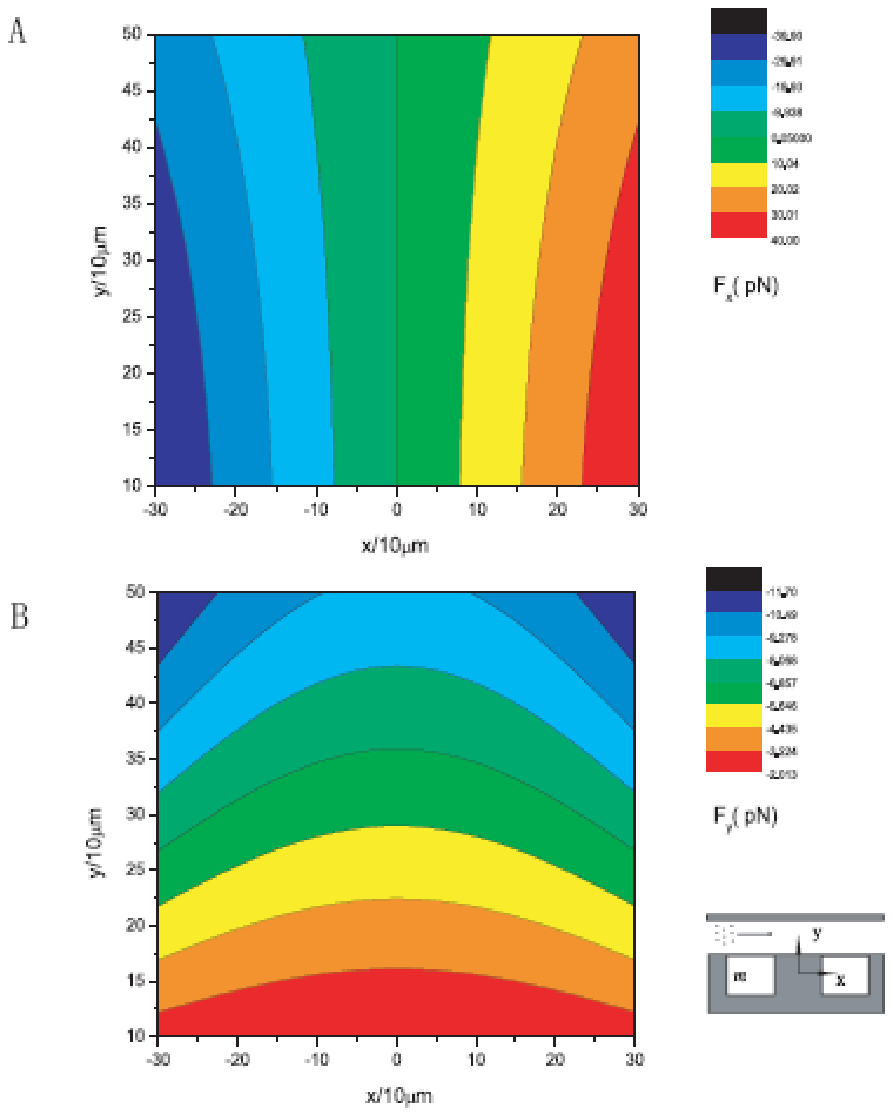}
\caption{Jian, Zhang and Huang}.\label{}
\end{figure}

\end{document}